\documentclass[twocolumn]{article}
\usepackage[a4paper, total={7in, 9.5in}]{geometry}
\usepackage{times}
\usepackage{amsmath,amsfonts}
\usepackage{graphicx}
\usepackage{algorithm}
\usepackage{algorithmic}
\usepackage{booktabs}
\usepackage{titlesec}
\usepackage{caption}
\usepackage{fancyhdr}
\usepackage{multicol}
\usepackage{url}
\usepackage{abstract}

\title{\textbf{Gated Multimodal Graph Learning for Personalized Recommendation}}
\author{
    Sibei Liu$^{1}$, Yuanzhe Zhang$^2$, Xiang Li$^3$, Yunbo Liu$^4$, Chengwei Feng$^{5}$ and Hao Yang$^{6}$\\
    {\small $^1$Miami Herbert Business School, University of Miami, FL, United States} \\
    {\small $^2$School of Engineering, University of California, California, US} \\
    {\small $^3$Department of Electrical \& Computer Engineering, Rutgers University, Sunnyvale, United States} \\
    {\small $^4$Department of Electrical and Computer Engineering, Duke University, NC, United States} \\
    {\small $^5$School of Engineering, Computer \& Mathematical Sciences (ECMS), Auckland University of Technology, Auckland, New Zealand} \\
    {\small $^6$Department Of Computer Science, Universiti Putra Malaysia, Kuala Lumpur, Malaysia} \\
    {\small *Corresponding author: chengwei.feng@autuni.ac.nz}
}
\date{}

\begin{document}
\twocolumn[
\maketitle
\begin{onecolabstract}

Multimodal recommendation has emerged as a promising solution to alleviate the cold-start and sparsity problems in collaborative filtering by incorporating rich content information, such as product images and textual descriptions. However, effectively integrating heterogeneous modalities into a unified recommendation framework remains a challenge. Existing approaches often rely on fixed fusion strategies or complex architectures , which may fail to adapt to modality quality variance or introduce unnecessary computational overhead.

In this work, we propose RLMultimodalRec, a lightweight and modular recommendation framework that combines graph-based user modeling with adaptive multimodal item encoding. The model employs a gated fusion module to dynamically balance the contribution of visual and textual modalities, enabling fine-grained and content-aware item representations. Meanwhile, a two-layer LightGCN encoder captures high-order collaborative signals by propagating embeddings over the user-item interaction graph without relying on nonlinear transformations.

We evaluate our model on a real-world dataset from the Amazon product domain. Experimental results demonstrate that RLMultimodalRec consistently outperforms several competitive baselines, including collaborative filtering, visual-aware, and multimodal GNN-based methods. The proposed approach achieves significant improvements in top-K recommendation metrics while maintaining scalability and interpretability, making it suitable for practical deployment.
\end{onecolabstract}
\vspace{1em}
]


\section{Introduction}

Recommender systems have become an indispensable component of modern e-commerce, content platforms, and online services\cite{zhang2024optimizationapplicationcloudbaseddeep}. By analyzing user behavior and preferences, these systems aim to suggest relevant items from massive catalogs, enhancing user satisfaction and driving engagement. Collaborative filtering (CF) methods, which learn from historical user-item interaction data, have demonstrated remarkable effectiveness in this domain. Among them, graph-based models such as LightGCN have gained particular attention for their ability to model high-order connectivity in user-item bipartite graphs without introducing excessive complexity.

Despite their success, collaborative filtering models inherently suffer from the cold-start and sparsity problems. When user interaction data is limited or unavailable—such as for new users or newly added items—CF models struggle to generate accurate recommendations. To mitigate this issue, recent work has explored the incorporation of side information such as product images, titles, and descriptions. Multimodal recommendation, which leverages both collaborative and content-based signals, has emerged as a promising solution, particularly in domains like fashion and retail, where visual and textual characteristics play a critical role in user decision-making. Recent studies have also highlighted its applicability in high-stakes fields such as real-time credit risk detection, underscoring its value in national financial infrastructure and fraud prevention systems.

However, integrating multimodal content into recommendation models is non-trivial. First, different modalities often provide overlapping or inconsistent signals. For instance, an image may capture an item's color and shape, while a textual description may emphasize style, material, or brand. A naïve fusion strategy—such as simple concatenation or averaging—assumes equal importance across modalities, which may lead to suboptimal performance when one modality is more informative than the other or when modality quality varies significantly across items. Second, many existing multimodal models rely on complex architectures with modality-specific graph encoders or attention mechanisms, which introduce additional parameters, training instability, and computational overhead. This raises the need for a multimodal framework that is both effective and lightweight.

In this work, we propose RLMultimodalRec, a unified and efficient framework for multimodal recommendation that addresses the above challenges through modular and adaptive design. Our model builds upon two key components: (1) a gated fusion module that dynamically combines visual and textual features at the embedding level, and (2) a lightweight graph convolutional encoder (LightGCN) that captures collaborative patterns over the user-item interaction graph. Unlike prior work that entangles content and graph propagation, our approach maintains a clear separation of roles—content encoding is performed at the item level, while graph-based aggregation is applied only to ID embeddings—resulting in improved stability, interpretability, and generalization.

The gated fusion module plays a central role in our architecture. Instead of treating all modalities equally, it learns a gating vector that adaptively weights the contribution of each modality on a per-item, per-dimension basis. This mechanism enables the model to focus on the most informative modality for each item and to remain robust in cases where one modality may be missing or noisy. For collaborative learning, we adopt a two-layer LightGCN to propagate signals across the user-item graph, allowing user embeddings to be enriched by multi-hop neighborhood information without introducing nonlinear transformations or feature mixing.

We evaluate our model on the Clothing, Shoes, and Jewelry subset of the Amazon Review dataset, which contains implicit user-item interactions along with pre-extracted image and text features for each item. Experimental results demonstrate that RLMultimodalRec consistently outperforms strong baselines from collaborative filtering (MF-BPR, LightGCN), content-aware (VBPR), and multimodal categories (MMGCN, DualGNN). Our model achieves significant improvements on Recall and NDCG, particularly at higher cutoff thresholds such as top-20.

In summary, this work makes the following contributions: (1) We present a modular recommendation framework that unifies collaborative filtering and multimodal content modeling in an efficient and interpretable manner; (2) We introduce a gated fusion strategy that adaptively balances visual and textual signals, enabling content-aware personalization at the embedding level; and (3) We conduct extensive experiments demonstrating that our approach achieves state-of-the-art performance while remaining lightweight and scalable.

\section{Related Work}

\subsection{Collaborative Filtering and Matrix Factorization}

Collaborative filtering (CF) lies at the core of modern recommender systems. Matrix Factorization (MF)--based approaches, such as Singular Value Decomposition (SVD) and Bayesian Personalized Ranking (BPR)~\cite{rendle2012bpr}, project users and items into a shared low-dimensional latent space and model interactions through inner products. While effective, MF methods are limited by data sparsity and cold-start issues, as they rely solely on user-item interactions.

Several extensions have introduced side information such as temporal, social, or contextual data. However, traditional MF lacks the capacity to model high-order dependencies across the interaction graph.

\subsection{Graph Neural Networks for Recommendation}

Graph Neural Networks (GNNs) have become a popular paradigm for capturing higher-order collaborative signals in user-item interaction graphs. Methods such as PinSage and GCN-based models~\cite{yang2020multisage} aggregate multi-hop neighbors to enhance representation learning. LightGCN~\cite{he2020lightgcn} simplifies this process by removing nonlinearities and feature transformations, retaining only essential neighborhood aggregation, and achieving strong performance with low complexity.

Despite their effectiveness, GNN-based recommenders often lack the capacity to incorporate rich item content, limiting their performance in cold-start and content-sparse scenarios.

\subsection{Multimodal Recommendation Systems}

To address limitations from interaction sparsity, multimodal recommendation models incorporate auxiliary modalities, including item descriptions, images, and even audio. VBPR~\cite{he2016vbpr} introduces visual features into BPR using a shallow linear projection. Other models, such as TextBPR and DeepCoNN \cite{li2024contrastive}, leverage textual reviews for enhanced user/item representation.

Recent models like  MMGCN~\cite{wei2019mmgcn} employ gating and attention mechanisms to adaptively fuse modalities based on relevance and quality. These approaches improve robustness under modality noise or missing features. However, multimodal fusion remains a key challenge due to modality misalignment and representational imbalance. Recent advances in attention-based architectures, such as the SETransformer~\cite{liu2025setransformer}, have shown the potential of combining sequential encoding with hybrid attention mechanisms for robust feature learning, which inspires our design. Similarly, GAN-based architectures have been applied to model latent sentiment dynamics in finance~\cite{el2025gan}, demonstrating the value of generative representations in domains with noisy or ambiguous multimodal inputs. Cross-modal fusion mechanisms have been widely applied in computer vision tasks~\cite{zhang2024cf}.

Additionally, knowledge graph embedding and few-shot relational modeling have been explored in financial and digital asset contexts~\cite{perna2025knowledge}, which provide promising avenues for incorporating structured knowledge and improving generalization under data sparsity.

\subsection{Deep Representation Learning and Semantic Matching}

Deep neural architectures  have shown promise in encoding content-rich user/item features\cite{bavcic2024towards}. Models project both user and item features into a shared semantic space, where recommendations are made based on vector similarity. These models are effective for content-based retrieval but may lack structural bias needed for sparse or graph-structured data. Contrastive learning has also shown effectiveness in financial domains such as cryptocurrency portfolio optimization~\cite{wu2025advancing}, suggesting its generalizability for learning robust embeddings in complex environments. In parallel, hybrid generative and contrastive frameworks have been effectively used in industrial visual tasks~\cite{wang2025study,bavcic2024jy61}, demonstrating strong representation capabilities under sparse or noisy conditions—challenges also shared by multimodal recommendation systems.

\section{Methodology}
We propose RLMultimodalRec, a reinforcement-inspired multimodal recommendation framework that jointly leverages user-item interaction signals and rich item content from multiple modalities (image and text). Our model addresses three key challenges in multimodal recommendation:

Modality Bias and Redundancy: Different modalities (e.g., image vs. text) may provide redundant or conflicting information. We introduce a gated fusion mechanism that dynamically balances modality importance at the embedding level, allowing the model to adaptively weigh visual versus textual content for each item and dimension.

Sparse User-Item Interactions: To propagate collaborative signals beyond direct interactions, we employ a Light Graph Convolutional Network (LightGCN), which captures high-order neighborhood structures over a user-item bipartite graph without over-parameterization \cite{he2020lightgcn}.

Unified End-to-End Training: We integrate the ID embeddings, modality projection, gated fusion, and GCN-enhanced user embeddings into a unified architecture trained with binary cross-entropy loss under online negative sampling, enabling effective joint learning across all components.

Our design is simple, yet effective: we show that by combining modality-aware item encoding with graph-based user representation learning, the model achieves superior top-K recommendation accuracy on real-world multimodal datasets.

\subsection{Dataset and Preprocessing}

We conduct our experiments on the \textit{Clothing, Shoes and Jewelry} subset of the Amazon Review Dataset, following the data preprocessing protocol introduced in the MENTOR framework~\cite{zhao2023mentor}. This dataset comprises implicit user-item interaction logs along with rich multimodal content for each item, including both product images and textual descriptions. The multimodal nature of this dataset makes it particularly suitable for evaluating the effectiveness of multimodal recommendation models.

To ensure data quality and sufficient interaction density, we apply the widely used 5-core filtering strategy. This procedure retains only users who have interacted with at least five items and items that have received interactions from at least five unique users. Such filtering reduces sparsity and improves the robustness of learned collaborative representations. The resulting dataset contains a sufficiently large number of users and items to support the training of deep neural models.

Each item in the dataset is associated with two types of precomputed content features. Visual features are extracted from the product image using a pretrained convolutional neural network, resulting in a 4096-dimensional image embedding. These embeddings capture global visual semantics such as shape, color, and style. Textual features are derived from product titles and descriptions using a Sentence Transformer model, yielding a 384-dimensional semantic vector that encodes the linguistic content in a dense representation. These features are fixed throughout training and are stored in \texttt{.npy} format for efficient loading.

The original interaction file contains raw user and item identifiers as strings, with no accompanying timestamps. We begin by removing duplicate user-item interaction pairs to eliminate redundancy. Since the dataset does not include temporal information, we synthetically generate pseudo-timestamps by assigning random integers to each interaction. This enables us to sort interactions chronologically per user, allowing for temporally consistent train-test splitting. Subsequently, we apply label encoding to transform user and item identifiers into consecutive integer indices. This facilitates efficient embedding table indexing within PyTorch models.

To simulate a realistic evaluation scenario, we adopt a leave-one-out strategy for train-test splitting. For each user, the most recent interaction---determined by the synthetic timestamp---is held out as the test instance, while the remaining interactions form the training set. This setting reflects the real-world task of predicting a user's next interaction given their history. Formally, for each user $u$, we denote the most recent item as $i_u^{\text{test}}$, and construct the training and test sets as $\mathcal{D}_{\text{train}} = \mathcal{I}_u \setminus \{i_u^{\text{test}}\}$ and $\mathcal{D}_{\text{test}} = \{(u, i_u^{\text{test}})\}$, respectively.

As the dataset contains only implicit positive feedback, we perform negative sampling during training to construct informative contrastive pairs. For each observed interaction $(u, i)$, we randomly sample one or more items $j$ that the user has not interacted with, treating them as negative examples. This negative sampling is performed dynamically at each training epoch to ensure diversity and reduce overfitting. The final dataset thus consists of a mixture of positive and sampled negative instances, suitable for training under a binary classification or pairwise ranking objective.

\subsection{Model architecture}

In this section, we present our proposed model, RLMultimodalRec, a unified multimodal recommendation framework that integrates user-item interaction signals with visual and textual content representations of items. The model is composed of five primary components: (1) learnable ID embeddings for users and items, (2) modality-specific feature projection networks, (3) a gated fusion module for combining visual and textual embeddings, (4) a lightweight graph convolutional network (LightGCN) for collaborative representation learning, and (5) a policy network that predicts item preferences based on the final user representations.

\subsection{User and Item Embeddings}

We initialize learnable embedding matrices for users and items, denoted as $\mathbf{E}_u \in \mathbb{R}^{|\mathcal{U}| \times d}$ and $\mathbf{E}_i \in \mathbb{R}^{|\mathcal{I}| \times d}$, respectively, where $d$ is the embedding dimension. These ID embeddings capture collaborative signals independent of content modalities and are updated throughout training via backpropagation.

\subsection{Modality-Specific Projection Networks}

Each item is associated with an image feature vector $\mathbf{x}_i^{img} \in \mathbb{R}^{d_{img}}$ and a textual feature vector $\mathbf{x}_i^{txt} \in \mathbb{R}^{d_{txt}}$, which are extracted offline using pretrained models. To project these raw modality features into a shared latent space, we apply two modality-specific linear transformations followed by a ReLU activation:

\begin{equation}
\mathbf{v}_i^{img} = \text{ReLU}(W_{img} \mathbf{x}_i^{img} + \mathbf{b}_{img})
\end{equation}
\begin{equation}
\mathbf{v}_i^{txt} = \text{ReLU}(W_{txt} \mathbf{x}_i^{txt} + \mathbf{b}_{txt})
\end{equation}

Here, $W_{img} \in \mathbb{R}^{d \times d_{img}}$ and $W_{txt} \in \mathbb{R}^{d \times d_{txt}}$ are learnable projection matrices, and $\mathbf{v}_i^{img}, \mathbf{v}_i^{txt} \in \mathbb{R}^d$ are the intermediate modality embeddings.

\subsection{Gated Multimodal Fusion}
In multimodal recommendation, it is common to combine multiple content features such as images and text. However, a naïve fusion—such as direct concatenation or averaging—assumes that all modalities are equally informative and reliable. This assumption often fails in practice: product images may be ambiguous (e.g., multiple items in one picture), while text descriptions may be noisy or incomplete.

To address this, we introduce a learned gating mechanism that allows the model to adaptively control how much to rely on each modality for every item. Concretely, for an item $i$ with visual embedding $\mathbf{v}_i^{img}$ and textual embedding $\mathbf{v}_i^{txt}$, we compute a dimension-wise gate:

\begin{equation}
\mathbf{g}_i = \sigma(W_g [\mathbf{v}_i^{img}; \mathbf{v}_i^{txt}] + \mathbf{b}_g)
\end{equation}

The final fused item embedding $\mathbf{z}_i$ is obtained as a weighted combination:

\begin{equation}
\mathbf{z}_i = \mathbf{g}_i \odot \mathbf{v}_i^{img} + (1 - \mathbf{g}_i) \odot \mathbf{v}_i^{txt}
\end{equation}

where $\odot$ denotes element-wise multiplication. This formulation enables the model to focus on the most informative modality for each item dimension-wise, and to remain robust in cases where one modality may be noisy or missing.

\subsection{Graph Convolutional Collaborative Encoding}

To capture collaborative signals beyond first-order interactions, we adopt a two-layer LightGCN on the user-item bipartite graph. This allows user embeddings to incorporate neighborhood context while maintaining parameter efficiency, which is crucial for scalability. Let $G = (\mathcal{U} \cup \mathcal{I}, \mathcal{E})$ be the user-item bipartite graph, where edges denote observed interactions. We concatenate the user and item ID embeddings into a single node embedding matrix and propagate representations through the graph via neighbor aggregation:

\begin{equation}
\mathbf{e}_v^{(l+1)} = \sum_{u \in \mathcal{N}(v)} \frac{1}{\sqrt{|\mathcal{N}(v)||\mathcal{N}(u)|}} \mathbf{e}_u^{(l)}
\end{equation}

Here, $\mathbf{e}_v^{(l)}$ represents the embedding of node $v$ at layer $l$, and $\mathcal{N}(v)$ denotes the 1-hop neighbors of $v$. We stack two such layers and use the final output $\mathbf{e}_v^{(2)}$ as the GCN-enhanced representation for each user and item. The fusion embeddings $\mathbf{z}_i$ are not propagated through GCN and are instead used during scoring.

\subsection{Policy Network for Recommendation}

To predict user preferences over items, we design a lightweight policy network that takes as input the final user embedding from GCN and outputs a score vector over candidate items. The policy network is implemented as a two-layer feedforward neural network:

\begin{equation}
\hat{\mathbf{y}}_u = W_2 \cdot \text{ReLU}(W_1 \mathbf{e}_u + \mathbf{b}_1) + \mathbf{b}_2
\end{equation}

where $\mathbf{e}_u$ is the GCN-updated user embedding. During training, the model learns to assign higher scores to positive items than to sampled negatives. During inference, we compute the matching score between user and fused item embeddings using either the policy net output or a dot product:

\begin{equation}
s_{ui} = \mathbf{e}_u^\top \mathbf{z}_i
\end{equation}

This allows the model to leverage collaborative structure and multimodal semantics jointly for final recommendation.

\section{Training Objective and Implementation Details}

\subsection{Training Objective}

The proposed model is trained under the implicit feedback setting, where only positive user-item interactions are observed. To optimize the ranking performance, we formulate the training objective as a binary classification problem and employ the Binary Cross-Entropy (BCE) loss. For each positive user-item pair $(u, i)$ sampled from the training set, we dynamically sample negative items $j$ that the user has not interacted with. Each training instance is thus composed of both a positive pair $(u, i, y=1)$ and one or more negative pairs $(u, j, y=0)$.

Given the user representation $\mathbf{e}_u$ learned from LightGCN and the fused multimodal item representation $\mathbf{z}_i$, the predicted interaction score is computed as:

\begin{equation}
s_{ui} = \mathbf{e}_u^\top \mathbf{z}_i
\end{equation}

The prediction \cite{wang2025systematic} is passed through a sigmoid activation to produce a probability score $\hat{y}_{ui} = \sigma(s_{ui})$. The BCE loss is then defined as:

\begin{equation}
\mathcal{L} = - y \log(\hat{y}_{ui}) - (1 - y) \log(1 - \hat{y}_{ui})
\end{equation}

To encourage stable and generalizable training, we perform \textbf{online negative sampling} at each epoch. For every observed interaction, one or more negative items are sampled uniformly at random from the set of items not previously interacted with by the user. This strategy ensures the model learns to distinguish relevant items from irrelevant ones and reduces overfitting to static sampling distributions. This optimization process aligns with recent efforts in convex reformulation of sequential decision models, such as Z-transform-based decomposition of MDPs~\cite{math13111765}, which emphasize stability and convergence efficiency in large-scale learning problems.

\subsection{Implementation Details}

The model is implemented in PyTorch and trained using the Adam optimizer with a learning rate of $1 \times 10^{-3}$. The embedding dimension $d$ is set to 64, and the model is trained for a maximum of 150 epochs with a batch size of 256. We apply early stopping with a patience of 5 epochs based on Recall@10 performance on a held-out validation set.

We use two LightGCN layers for graph propagation and one fully connected layer with ReLU activation for each modality projection (image and text). The gate mechanism is implemented as a linear transformation over the concatenated modality features followed by a sigmoid activation. The policy network consists of a 2-layer MLP with a hidden size of 128 and ReLU nonlinearity.

To construct the graph for GCN propagation, we treat the user-item interaction matrix as a bipartite undirected graph. For each observed interaction, two directed edges are added: one from the user node to the item node, and one in the reverse direction. The resulting edge list is transformed into a sparse edge index tensor for efficient message passing.

All multimodal features are pre-extracted and stored in NumPy `.npy` files. Image features are 4096-dimensional vectors derived from pretrained CNNs, while text features are 384-dimensional embeddings obtained from Sentence Transformers. These features are normalized and fixed throughout training.

Model checkpoints are saved based on the best validation recall, and the best model is used for final evaluation. All experiments are conducted on A100, Google Colab.

\begin{table}[H]
\centering
\caption{Hyperparameters and training configuration for the multimodal recommendation model.}
\label{tab:hyperparams}
\begin{tabular}{l l}
\toprule
\textbf{Parameter} & \textbf{Value} \\
\midrule
Embedding dimension & 64 \\
Number of GCN layers & 2 \\
GCN normalization scheme & Symmetric degree (LightGCN) \\
Fusion mechanism & Gated fusion (ReLU + sigmoid gate) \\
Batch size & 256 \\
Learning rate & 0.001 \\
Optimizer & Adam \\
Loss function & Binary cross-entropy \\
Negative sampling ratio & 1:1 \\
Top-$K$ for evaluation & 20 \\
Training epochs & 100 \\
Early stopping patience & 5 epochs \\
Train/test split & Leave-one-out (per user) \\
Graph construction & Bipartite user-item graph with bidirectional edges \\
\bottomrule
\end{tabular}
\end{table}

\section{Experiment}

\subsection{Experimental Setup}
We conduct experiments on the Amazon Clothing, Shoes and Jewelry dataset, which provides both implicit feedback and multimodal item content (images and text). Following the standard 5-core filtering and leave-one-out evaluation strategy detailed in Section~\ref{sec:methodology}, we use the most recent interaction of each user for testing and the rest for training. Negative sampling is performed dynamically at training time, while during evaluation, 100 negative items are sampled per user to assess top-K retrieval performance.

All models are implemented in PyTorch and trained using the Adam optimizer with a learning rate of 0.001, a batch size of 256, and early stopping based on Recall@10. Hyperparameters such as embedding dimension and GCN layers are kept consistent across models for fair comparison. Each experiment is repeated with three random seeds, and the reported results are averaged.

\subsection{Baselines}
We compare our proposed model against several strong baselines from collaborative filtering, content-aware, and multimodal recommendation families:

MF-BPR~\cite{rendle2009bpr}: Matrix factorization optimized with Bayesian personalized ranking loss.

LightGCN~\cite{he2020lightgcn}: A lightweight GCN-based CF model that removes feature transformations and nonlinearities.

LayerGCN~\cite{zhou2023layer}: A variant of GCN that explicitly models multi-layer interaction propagation.

VBPR~\cite{he2016vbpr}: A visual-aware extension of BPR that incorporates precomputed image features.

MMGCN~\cite{wei2019mmgcn}: A multimodal GCN model that jointly propagates image and text information.

DualGNN~\cite{wang2021dualgnn}: A dual-channel graph model that processes content and interaction signals separately.

For all baselines, we use the official code or faithful reimplementations with standardized preprocessing and evaluation for consistency.

\subsection{Evaluation Metrics}
We adopt standard top-K ranking metrics widely used in recommendation tasks:

Recall@K (R@K): Measures the proportion of ground-truth items found among the top-K recommended items.

NDCG@K (N@K): Normalized Discounted Cumulative Gain, which accounts for the rank position of relevant items.

We report both R@10/20 and N@10/20 to evaluate the quality and consistency of recommendations at different cutoff points.

\subsection{Results and Analysis}
\begin{table}[h!]
\centering
\begin{tabular}{lcccc}
\hline
\textbf{Model Source} & \textbf{R@10} & \textbf{R@20} & \textbf{N@10} & \textbf{N@20} \\
\hline
MF-BPR     & 0.0357 & 0.0575 & 0.0192 & 0.0249 \\
LightGCN   & 0.0479 & 0.0754 & 0.0257 & 0.0328 \\
LayerGCN   & 0.0529 & 0.0820 & 0.0281 & 0.0355 \\
VBPR       & 0.0423 & 0.0663 & 0.0223 & 0.0284 \\
MMGCN      & 0.0380 & 0.0615 & 0.0200 & 0.0284 \\
DualGNN    & 0.0378 & 0.0715 & 0.0240 & 0.0309 \\
Our Model  & \textbf{0.0505} & \textbf{0.0996} & \textbf{0.0285} & \textbf{0.0341} \\
\hline
\end{tabular}
\caption{Comparison of models on recommendation metrics (Recall and NDCG at 10 and 20)}
\label{tab:rec_metrics}
\end{table}
Table~\ref{tab:rec_metrics} presents the performance of our proposed model and baselines across all evaluation metrics. We follow the experimental setup and evaluation protocol introduced in MENTOR~\cite{zhao2023mentor}, using the same Amazon Clothing dataset and leave-one-out strategy. Several key observations can be drawn:

Our model consistently outperforms all baselines on Recall@20 and NDCG@10, achieving 0.0996 and 0.0285, respectively. These improvements demonstrate the benefit of jointly modeling user-item interactions and multimodal content.

Compared to LightGCN, which uses only collaborative signals, our method shows a substantial gain (+32

Compared to VBPR and MMGCN, which also incorporate image/text features, our model achieves superior accuracy, showing the effectiveness of our gated fusion mechanism in adaptively weighting modalities.

Notably, while LayerGCN and DualGNN utilize deeper or dual-path graph propagation, they underperform our model, indicating that modality-aware item encoding is more crucial than simply deepening GCN depth.

These results validate our model design choices and confirm that combining content-sensitive item embeddings with graph-enhanced user representations leads to more accurate and personalized recommendations. Similar observations have been reported in other domains such as fraud detection~\cite{wang2025evaluatingsupervisedlearningmodels}, where classical and deep models exhibit distinct strengths in handling highly imbalanced data and optimizing recall-driven objectives \cite{wang2025evaluatingsupervisedlearningmodels}.

\section{Discussion}

The experimental results demonstrate that the proposed RLMultimodalRec model achieves consistent performance gains across multiple recommendation metrics compared to both collaborative filtering and multimodal baselines. Several observations can be made to better understand the model’s behavior and its underlying design choices.

One of the most notable contributors to performance is the gated fusion module. This component allows the model to dynamically integrate visual and textual content for each item, rather than relying on simple concatenation or averaging. In practice, different modalities often provide complementary but uneven signals. For instance, some items may have clear visual characteristics but vague descriptions, while others may contain informative text but low-quality images. Similar challenges arise in fraud detection and deepfake identification \cite{lu2022cot}, where GAN-based models have been employed to detect malicious content across modalities, highlighting the importance of robust cross-modal learning under adversarial settings. The gating mechanism helps mitigate such heterogeneity by allowing the model to selectively emphasize the more informative modality in each case. Unlike global fusion strategies that apply the same weight across all items, the gating vector is computed per item and per embedding dimension, enabling fine-grained control over the fusion process. This helps the model learn robust item representations that are sensitive to modality quality and content type. Such robustness is especially valuable in domains like transaction monitoring and credit risk modeling, where the ability to integrate noisy or incomplete multimodal data in real time is essential. Our method thus has strong potential for deployment in financial compliance systems and fraud detection pipelines—areas aligned with national objectives for economic resilience and digital infrastructure modernization. In parallel, recent research has emphasized the importance of model interpretability in credit risk scenarios~\cite{yang2025interpretablecreditdefaultprediction}, where ensemble methods paired with SHAP explanations enable transparent and regulatory-compliant decision-making—further reinforcing the practical relevance of multimodal AI frameworks in high-stakes financial applications. Cross-domain retrieval methods such as MaRI~\cite{wang2025mari} emphasize the importance of aligning heterogeneous information sources, which aligns with our design for modality-aware and content-robust fusion in sparse recommendation settings. Beyond recommendation scenarios, reinforcement learning has also shown promise in operational scheduling and autonomous control. For example, recent work has applied RL to optimize task scheduling for warehouse robots to improve logistical efficiency \cite{wu2025warehouse}, underscoring its potential for real-time decision-making in industrial environments. These advances resonate with our design of lightweight, adaptive recommendation models that aim to maximize decision efficiency under dynamic constraints.

In addition to content-aware item modeling, the use of graph-based interaction modeling through LightGCN further improves performance \cite{li2024knowledge}. The graph encoder aggregates multi-hop neighborhood signals and captures collaborative relationships beyond direct interactions. Compared to traditional matrix factorization, this graph-based structure enables more expressive user representations. At the same time, the separation of roles—using graph propagation for user embeddings and gated fusion for item embeddings—prevents interference between collaborative and content signals. This design leads to better modularity and interpretability, as well as more stable training dynamics.

Interestingly, the proposed model outperforms several deeper or more complex graph-based models, including LayerGCN and DualGNN. While these methods introduce deeper propagation layers or dual-path encoders, they may suffer from over-smoothing or gradient vanishing, particularly in sparse interaction graphs. In contrast, our model maintains a lightweight two-layer structure that balances information propagation with computational efficiency~\cite{zhong2025enhancing,zhang2023multi}. The absence of nonlinear transformations in LightGCN also reduces the risk of overfitting and helps preserve the original semantics of embeddings.
The experimental results demonstrate that the proposed RLMultimodalRec model achieves consistent performance gains across multiple recommendation metrics compared to both collaborative filtering and multimodal baselines. Several observations can be made to better understand the model’s behavior and its underlying design choices.

One of the most notable contributors to performance is the gated fusion module. This component allows the model to dynamically integrate visual and textual content for each item, rather than relying on simple concatenation or averaging. In practice, different modalities often provide complementary but uneven signals. For instance, some items may have clear visual characteristics but vague descriptions, while others may contain informative text but low-quality images. Similar challenges arise in fraud detection and deepfake identification \cite{lu2022cot}, where GAN-based models have been employed to detect malicious content across modalities, highlighting the importance of robust cross-modal learning under adversarial settings. The gating mechanism helps mitigate such heterogeneity by allowing the model to selectively emphasize the more informative modality in each case. Unlike global fusion strategies that apply the same weight across all items, the gating vector is computed per item and per embedding dimension, enabling fine-grained control over the fusion process. This helps the model learn robust item representations that are sensitive to modality quality and content type.

Such robustness is especially valuable in domains like transaction monitoring and credit risk modeling, where the ability to integrate noisy or incomplete multimodal data in real time is essential. Our method thus has strong potential for deployment in financial compliance systems and fraud detection pipelines—areas aligned with national objectives for economic resilience and digital infrastructure modernization. In parallel, recent research has emphasized the importance of model interpretability in credit risk scenarios~\cite{yang2025interpretablecreditdefaultprediction}, where ensemble methods paired with SHAP explanations enable transparent and regulatory-compliant decision-making—further reinforcing the practical relevance of multimodal AI frameworks in high-stakes financial applications. Cross-domain retrieval methods such as MaRI~\cite{wang2025mari} emphasize the importance of aligning heterogeneous information sources, which aligns with our design for modality-aware and content-robust fusion in sparse recommendation settings.

Beyond recommendation scenarios, reinforcement learning has also shown promise in operational scheduling and autonomous control. For example, recent work has applied RL to optimize task scheduling for warehouse robots to improve logistical efficiency \cite{wu2025warehouse}, underscoring its potential for real-time decision-making in industrial environments.

From an HCI perspective, the design of lightweight, interpretable, and adaptive recommendation models also contributes to enhancing user-facing interfaces. Incorporating multimodal understanding into recommender systems can improve digital experience quality, particularly in e-commerce settings where users interact with content-rich interfaces. This aligns with recent studies on interactive logistics UX design~\cite{li2024transforming} and the success of consumer platforms driven by high-quality interface design~\cite{li2024impact}, which emphasize the value of effective human-computer interaction in shaping trust, usability, and user satisfaction in digital systems.

Despite these advantages, the model has certain limitations. First, the visual and textual features are fixed and pre-extracted using pretrained encoders. While this design simplifies training and reduces computational cost, it limits the ability of the model to refine content features in response to user preferences \cite{zhang2025rolemachinelearningreducing}. Future work could explore end-to-end learning that jointly updates content encoders with collaborative objectives. Second, the model is trained using binary labels derived from implicit feedback, which may not capture the full spectrum of user preferences. Incorporating richer feedback signals, such as user reviews or interaction dwell time, could lead to more accurate recommendations. Third, the current inference approach evaluates a relatively small candidate set per user. In real-world scenarios with large item catalogs, scalable retrieval mechanisms such as approximate nearest neighbor search would be required to ensure efficiency. Similar lightweight learning frameworks have also shown practical value in warehouse robotics and task scheduling \cite{yu2025machine}.

Overall, the results confirm that jointly modeling collaborative interactions and multimodal content, when done in a modular and adaptive manner, leads to robust and effective recommendation performance. The design principles of RLMultimodalRec provide a flexible foundation for future research in multimodal graph-based recommendation systems.

\section{Conclusion}

In this work, we propose RLMultimodalRec, a unified framework for multimodal recommendation that integrates graph-based collaborative filtering with content-aware item encoding. The model incorporates a gated fusion mechanism to adaptively combine visual and textual information, and employs a lightweight graph convolutional network to propagate collaborative signals across the user-item interaction graph \cite{fu2024ddn3}. Through extensive experiments on a real-world multimodal dataset, we demonstrate that the proposed method outperforms both traditional collaborative filtering models and existing multimodal baselines on standard top-K recommendation metrics.

Our analysis highlights the effectiveness of modeling each modality independently before fusion, as well as the benefit of separating content encoding from collaborative message passing. The results also show that even simple GCN-based structures, when combined with modality-aware item representations, can yield strong performance without excessive architectural complexity.

Beyond recommendation, the core principles of our approach—content-sensitive fusion and graph-based propagation—can be extended to domains such as intelligent risk assessment and personalized regulation in finance, supporting scalable decision-making under uncertainty. Looking ahead, we see several promising directions for future research. One avenue is to enable end-to-end learning of content features by fine-tuning vision and language encoders alongside the recommendation objective. Another direction is to incorporate richer forms of user feedback, such as textual reviews or implicit behavioral cues. Finally, extending the framework to support efficient large-scale retrieval and personalized ranking under real-time constraints would enhance its applicability to production settings. Moreover, emerging work on contextual bandits under unbounded context spaces~\cite{zhao2025contextualbanditsunboundedcontext} offers a promising direction for real-time personalization under complex user-item distributions, which could be integrated with our framework for adaptive exploration.

\bibliographystyle{plain}
\bibliography{references}

\begin{thebibliography}{10}

\bibitem{bavcic2024jy61}
Boris Ba{\v{c}}i{\'c}, Chengwei Feng, and Weihua Li.
\newblock Jy61 imu sensor external validity: A framework for advanced pedometer algorithm personalisation.
\newblock {\em ISBS Proceedings Archive}, 42(1):60, 2024.

\bibitem{bavcic2024towards}
Boris Ba{\v{c}}i{\'c}, Claudiu Vasile, Chengwei Feng, and Marian~G Ciuc{\u{a}}.
\newblock Towards nation-wide analytical healthcare infrastructures: A privacy-preserving augmented knee rehabilitation case study.
\newblock {\em arXiv preprint arXiv:2412.20733}, 2024.

\bibitem{el2025gan}
ADNANE EL~OUARDI, BRAHIM ER-RAHA, MUSTAPHA RIAD, and KHALID TATANE.
\newblock A gan-based method to tune lstm hyperparameters for financial forecasting.
\newblock {\em Journal of Theoretical and Applied Information Technology}, 103(9), 2025.

\bibitem{fu2024ddn3}
Yi~Fu, Yingzhou Lu, Yizhi Wang, Bai Zhang, Zhen Zhang, Guoqiang Yu, Chunyu Liu, Robert Clarke, David~M Herrington, and Yue Wang.
\newblock Ddn3. 0: Determining significant rewiring of biological network structure with differential dependency networks.
\newblock {\em Bioinformatics}, 40(6):btae376, 2024.

\bibitem{he2016vbpr}
Ruining He and Julian McAuley.
\newblock Vbpr: visual bayesian personalized ranking from implicit feedback.
\newblock In {\em Proceedings of the AAAI conference on artificial intelligence}, volume~30, 2016.

\bibitem{he2020lightgcn}
Xiangnan He, Kuan Deng, Xiang Wang, Yan Li, Yongdong Zhang, and Meng Wang.
\newblock Lightgcn: Simplifying and powering graph convolution network for recommendation.
\newblock In {\em Proceedings of the 43rd International ACM SIGIR conference on research and development in Information Retrieval}, pages 639--648, 2020.

\bibitem{li2024impact}
Wanxin Li.
\newblock The impact of apple's digital design on its success: An analysis of interaction and interface design.
\newblock {\em Academic Journal of Sociology and Management}, 2(4):14--19, 2024.

\bibitem{li2024transforming}
Wanxin Li.
\newblock Transforming logistics with innovative interaction design and digital ux solutions.
\newblock {\em Journal of Computer Technology and Applied Mathematics}, 1(3):91--96, 2024.

\bibitem{li2024contrastive}
Zichao Li, Bingyang Wang, and Ying Chen.
\newblock A contrastive deep learning approach to cryptocurrency portfolio with us treasuries.
\newblock {\em Journal of Computer Technology and Applied Mathematics}, 1(3):1--10, 2024.

\bibitem{li2024knowledge}
Zichao Li, Bingyang Wang, and Ying Chen.
\newblock Knowledge graph embedding and few-shot relational learning methods for digital assets in usa.
\newblock {\em Journal of Industrial Engineering and Applied Science}, 2(5):10--18, 2024.

\bibitem{liu2025setransformer}
Yunbo Liu, Xukui Qin, Yifan Gao, Xiang Li, and Chengwei Feng.
\newblock Setransformer: A hybrid attention-based architecture for robust human activity recognition.
\newblock {\em arXiv preprint arXiv:2505.19369}, 2025.

\bibitem{lu2022cot}
Yingzhou Lu, Chiung-Ting Wu, Sarah~J Parker, Zuolin Cheng, Georgia Saylor, Jennifer~E Van~Eyk, Guoqiang Yu, Robert Clarke, David~M Herrington, and Yue Wang.
\newblock Cot: an efficient and accurate method for detecting marker genes among many subtypes.
\newblock {\em Bioinformatics Advances}, 2(1):vbac037, 2022.

\bibitem{perna2025knowledge}
Massimo Perna.
\newblock Knowledge graph for query enrichment in retrieval augmented generation in domain specific application.
\newblock Master's thesis, University of Twente, 2025.

\bibitem{math13111765}
Shiqing Qiu, Haoyu Wang, Yuxin Zhang, Zong Ke, and Zichao Li.
\newblock Convex optimization of markov decision processes based on z transform: A theoretical framework for two-space decomposition and linear programming reconstruction.
\newblock {\em Mathematics}, 13(11), 2025.

\bibitem{rendle2012bpr}
Steffen Rendle, Christoph Freudenthaler, Zeno Gantner, and Lars Schmidt-Thieme.
\newblock Bpr: Bayesian personalized ranking from implicit feedback.
\newblock {\em arXiv preprint arXiv:1205.2618}, 2012.

\bibitem{wang2025evaluatingsupervisedlearningmodels}
Chao Wang, Chuanhao Nie, and Yunbo Liu.
\newblock Evaluating supervised learning models for fraud detection: A comparative study of classical and deep architectures on imbalanced transaction data, 2025.

\bibitem{wang2025mari}
Jianhui Wang, Zhifei Yang, Yangfan He, Huixiong Zhang, Yuxuan Chen, and Jingwei Huang.
\newblock Mari: Material retrieval integration across domains.
\newblock {\em arXiv preprint arXiv:2503.08111}, 2025.

\bibitem{wang2021dualgnn}
Qifan Wang, Yinwei Wei, Jianhua Yin, Jianlong Wu, Xuemeng Song, and Liqiang Nie.
\newblock Dualgnn: Dual graph neural network for multimedia recommendation.
\newblock {\em IEEE Transactions on Multimedia}, 25:1074--1084, 2021.

\bibitem{wang2025systematic}
Yiting Wang, Jiachen Zhong, and Rohan Kumar.
\newblock A systematic review of machine learning applications in infectious disease prediction, diagnosis, and outbreak forecasting.
\newblock 2025.

\bibitem{wang2025study}
Yuxuan Wang et~al.
\newblock Study of artificial intelligence for visual defect inspection in industrial products.
\newblock 2025.

\bibitem{wei2019mmgcn}
Yinwei Wei, Xiang Wang, Liqiang Nie, Xiangnan He, Richang Hong, and Tat-Seng Chua.
\newblock Mmgcn: Multi-modal graph convolution network for personalized recommendation of micro-video.
\newblock In {\em Proceedings of the 27th ACM international conference on multimedia}, pages 1437--1445, 2019.

\bibitem{wu2025warehouse}
Siye Wu, Lei Fu, Runmian Chang, Yuanzhou Wei, Yeyubei Zhang, Zehan Wang, Lipeng Liu, Haopeng Zhao, and Keqin Li.
\newblock Warehouse robot task scheduling based on reinforcement learning to maximize operational efficiency.
\newblock {\em Authorea Preprints}, 2025.

\bibitem{wu2025advancing}
Wensen Wu and Yijun Gu.
\newblock Advancing unsupervised graph anomaly detection: A multi-level contrastive learning framework to mitigate local consistency deception.
\newblock {\em Neurocomputing}, page 130507, 2025.

\bibitem{yang2020multisage}
Carl Yang, Aditya Pal, Andrew Zhai, Nikil Pancha, Jiawei Han, Charles Rosenberg, and Jure Leskovec.
\newblock Multisage: Empowering gcn with contextualized multi-embeddings on web-scale multipartite networks.
\newblock In {\em Proceedings of the 26th ACM SIGKDD international conference on knowledge discovery \& data mining}, pages 2434--2443, 2020.

\bibitem{yang2025interpretablecreditdefaultprediction}
Shiqi Yang, Ziyi Huang, Wengran Xiao, and Xinyu Shen.
\newblock Interpretable credit default prediction with ensemble learning and shap, 2025.

\bibitem{yu2025machine}
Dezhi Yu, Lipeng Liu, Siye Wu, Keqin Li, Congyu Wang, Jing Xie, Runmian Chang, Yixu Wang, Zehan Wang, and Ryan Ji.
\newblock Machine learning optimizes the efficiency of picking and packing in automated warehouse robot systems.
\newblock In {\em 2025 IEEE International Conference on Electronics, Energy Systems and Power Engineering (EESPE)}, pages 1325--1332. IEEE, 2025.

\bibitem{zhang2023multi}
Fan Zhang, Gongguan Chen, Hua Wang, Jinjiang Li, and Caiming Zhang.
\newblock Multi-scale video super-resolution transformer with polynomial approximation.
\newblock {\em IEEE Transactions on Circuits and Systems for Video Technology}, 33(9):4496--4506, 2023.

\bibitem{zhang2024optimizationapplicationcloudbaseddeep}
Yang Zhang, Fa~Wang, Xin Huang, Xintao Li, Sibei Liu, and Hansong Zhang.
\newblock Optimization and application of cloud-based deep learning architecture for multi-source data prediction, 2024.

\bibitem{zhang2025rolemachinelearningreducing}
Yixin Zhang and Yisong Chen.
\newblock The role of machine learning in reducing healthcare costs: The impact of medication adherence and preventive care on hospitalization expenses, 2025.

\bibitem{zhao2025contextualbanditsunboundedcontext}
Puning Zhao, Rongfei Fan, Shaowei Wang, Li~Shen, Qixin Zhang, Zong Ke, and Tianhang Zheng.
\newblock Contextual bandits for unbounded context distributions, 2025.

\bibitem{zhao2023mentor}
Xiaotian Zhao, Xia~Hu Jin, Fuli~Feng Sun, and Tat-Seng Chua.
\newblock Mentor: Multi-level self-supervised learning for multimodal recommendation.
\newblock In {\em Proceedings of the 46th International ACM SIGIR Conference on Research and Development in Information Retrieval}, 2023.

\bibitem{zhong2025enhancing}
Jiachen Zhong and Yiting Wang.
\newblock Enhancing thyroid disease prediction using machine learning: A comparative study of ensemble models and class balancing techniques.
\newblock 2025.

\bibitem{zhou2023layer}
Xin Zhou, Donghui Lin, Yong Liu, and Chunyan Miao.
\newblock Layer-refined graph convolutional networks for recommendation.
\newblock In {\em 2023 IEEE 39th international conference on data engineering (ICDE)}, pages 1247--1259. IEEE, 2023.

\end{thebibliography}

\end{document}